# Differential orientation and conformation of Keratinocyte Growth Factor observed at HEMA, HEMA/MMA, and HEMA/MAA hydrogel surfaces developed for wound healing

Shohini Sen-Britain[a]

State University of New York at Buffalo, Department of Chemistry, 475 Natural Sciences Complex, Buffalo, NY 14221, USA

Wesley Hicks[b]

Roswell Comprehensive Cancer Center, Department of Head and Neck/Plastic and Reconstructive Surgery, 665 Elm Street, Buffalo, NY 14203, USA

Robert Hard[c]

State University of New York at Buffalo, Jacobs School of Medicine and Biomedical Sciences, Department of Pathological and Anatomical Sciences, 955 Main St, Buffalo, NY 14203, USA

Joseph A. Gardella Jr.[d]

State University of New York at Buffalo, Department of Chemistry, 475 Natural Sciences Complex, Buffalo, NY 14221, USA

[d]American Vacuum Society member

[d]Electronic mail: gardella@buffalo.edu

**Abstract.** The development of hydrogels for protein delivery requires protein-hydrogel interactions that cause minimal disruption of the protein's biological activity. Biological activity can be influenced by factors such as orientation and conformation. Hydrogels must promote the adsorption of biomolecules onto the surface and the diffusion of biomolecules into the porous network at the surface, while maintaining native protein conformation, keeping the protein in an accessible orientation for receptor binding, and maximizing protein release. We report here the evaluation of (hydroxyethyl) methacrylate (HEMA)-based hydrogel systems for the delivery of keratinocyte growth factor (KGF) to promote re-epithelialization in wound healing. In this work, we characterize two hydrogel blends in addition to HEMA alone, and report how protein orientation, conformation, and protein release is affected. The first blend incorporates methyl methacrylate (MMA), which is known to promote adsorption of protein to its surface due to its hydrophobicity. The second blend incorporates methacrylic acid (MAA), which is known to promote the diffusion of protein into its surface due to its hydrophilicity. We find that the KGF at the surface of the HEMA/MMA blend appears to be more orientationally accessible and conformationally active than KGF at the surface of the HEMA/MAA blend. We also report that KGF at the surface of the HEMA/MAA blend becomes conformationally denatured, likely due to hydrogen bonding. While KGF at the surface of these blends can be differentiated by FTIR-ATR spectroscopy and ToF-SIMS in conjunction with PCA, KGF swelling, uptake, and release profiles are indistinguishable. The differences in KGF orientation and conformation between these blends may result in different biological responses in future cell-based experiments.

# I. INTRODUCTION

The development of hydrogels as clinically translatable protein delivery devices involves engineering a delicate balance of efficient, localized and controlled protein release, while retaining the native conformation and orientation of protein in order for it to perform its biological function. (Hydroxyethyl)methacrylate (HEMA) hydrogels have been of interest for several decades as drug delivery vehicles since adsorption and diffusion of biomolecules into their porous network structure was observed in contact lens research[1]. HEMA hydrogels are swelling materials that are able to take up a large amount of water which gives them properties similar to those of biological tissues[2]. While the development of HEMA or alternative hydrogel-based small molecule delivery has been extensively studied given the ease of diffusion into the hydrogel porous network, protein delivery poses more complex issues[3]. The hydrogel must possess characteristics that efficiently promote adsorption and diffusion of protein onto the surface and into the porous network, (1) while maintaining the native conformation of the protein, (2) presenting the protein in an orientation allowing for binding to its target cell surface receptors, and (3) allowing for maximized protein release.

We are specifically interested in the development of a bioadhesive protein delivery device to treat traumatic or chronic wounds of the epithelium in regions such as the skin, eyes, throat etc. Upon traumatic injury, the epithelium initiates an intra-epithelial repair process which involves the expression of mitogenic and motogenic proteins such as keratinocyte growth factor (KGF)[4]. Prior work from our labs has shown that exogenous addition of KGF in *in vitro* wound models can expedite the wound closure process[5]. Biomaterial-based KGF delivery allows for localized and controlled delivery of KGF to the wound[4, 6-9]. The goal of our present research program is to develop and characterize a HEMA-based KGF delivery system that is to be used as an interface between the connective tissue and epithelium, while stimulating the wound healing process. The wound healing process is initiated by the binding of KGF to the KGF-receptor, which is a cell surface signal transducing receptor of the tyrosine kinase family[10]. This binding event leads to cell division and migration of epithelial cells to the wound site, thereby stimulating the wound healing process. KGF binding to the KGF-receptor is mediated by KGF binding to heparin prior to receptor binding, which eventually results in the formation of a KGF-heparin-KGF receptor complex. Therefore, the heparin binding ability of KGF is an initial indicator of its biological activity, and KGF must retain its native conformation and orientation in order to carry out this role[10].

Previous studies utilizing 2D correlation spectroscopy of a HEMA-based KGF delivery system developed in our labs have shown that while the HEMA system allows for efficient loading and extended release of KGF, 40% of the loaded KGF becomes trapped at the hydrogel surface and denatures. The results of this study showed that the interaction between HEMA and KGF disrupts loop structures of the protein responsible for heparin binding interactions. Therefore, we suspect that the interaction between HEMA and KGF mimics the KGF-heparin interaction due to the hydrophilic nature of HEMA[11].

In order to evaluate modified, second generation systems for KGF delivery that move towards circumventing the issues of incomplete release and eventual denaturation, we have chosen to pull together knowledge from prior work on protein adsorption on biomaterial surfaces and prior work on hydrogel development.

Protein adsorption on biomaterial surfaces has been viewed as an issue to understand and prevent; certain materials of interest as biomedical implants have been shown to cause adsorption of common blood proteins[12]. These blood proteins such as serum albumin, fibrinogen, fibronectin, and von Willebrand factor adsorb in orientations that allow for binding to cell surface receptors which leads to the activation of pathways involved in clotting and potential thrombus formation[12-26]. While the activation of thrombus formation by blood proteins due to adsorption onto implant surfaces is not ideal and potentially catastrophic, we want to promote KGF orientations at our hydrogel surface that are able to bind heparin and the KGF cell surface receptor. However, surfaces of interest as implants such as polystyrene, polycarbonate, titanium, poly-methyl methacryate etc. rarely possess characteristics of a drug delivery device such as the ability to load proteins into a material, and release and deliver proteins over time. It is possible that a blend of a material promoting adsorption with HEMA may lead to a receptor-accessible orientation of KGF at the surface.

Hydrogels are known to present viscoelastic features similar to those of biological tissues. Pores within the hydrogel network have the capacity to incorporate components from different biological matrices, and release of these components can be adjusted by varying pore size created by the percent of crosslinker used. Hydrogels also are reported to have good biocompatibility given that no pH changes occur during release due to degradation, and because organic solvents are not required for synthesis. Release of protein has been reported to be due to the collapse of pores in HEMA hydrogels, and the controlled release of numerous proteins has been investigated[3, 15, 27]. HEMA hydrogels also have bioadhesive properties, and are currently FDA approved wound adhesives[28].

However, hydrogel based protein delivery systems have yet to be clinically translated due to the lack of studies on the biological activity of released proteins and proteins at the hydrogel surface throughout the release process. Hydrophilicity of the hydrogel promotes diffusion of water and solutes into the porous network. Carboxylic acid-containing monomers such as methacrylic acid have been reported to increase hydrophilicity[29-33]. A HEMA/MAA hydrogel may promote increased diffusion of KGF and potentially increase the concentration of KGF in and at the hydrogel surface. However, this may be negated by stronger interactions between KGF and the hydrogel, which could lead to poor release profiles and potential denaturation of KGF at the hydrogel surface.

Our approach to the development and evaluation of second-generation KGF delivery systems is to synthesize hydrogel blends that incorporate (1) methyl methacrylate (MMA), which is known to promote protein adsorption, and (2) methacrylic acid (MAA), which promotes a higher hydrogel water content and increased protein diffusion[34]. For the reasons we have argued above, MMA and MAA containing HEMA hydrogels were also extensively studied in contact lens applications in studies on preventing protein adsorption[27, 35-39]. However, to the best of our knowledge, the surface characterization of these blends using ToF-SIMS has not been reported. We report the characterization of these hydrogels, FTIR-ATR spectroscopy for the analysis of KGF conformation, ToF-SIMS in conjunction with principal component analysis (PCA) to compare KGF orientation across the surface of the HEMA/MAA, HEMA/MMA, and HEMA blends, as well as release profiles for these hydrogels.

## II. EXPERIMENTAL

2-hydroxyethyl methacrylate, (HEMA, contains ≤ 50 ppm monomethyl ether hydroquinone as inhibitor, SKU 477028), methacrylic acid (MAA, contains 250 ppm monomethyl ether hydroquinone as inhibitor, SKU 155721), methyl methacrylate (MMA, ≤ 30 ppm monomethyl ether hydroquinone as inhibitor, SKU 55909), trimethylolpropane triacrylate (TMPTMA, contains 600 ppm monomethyl ether hydroquinone as inhibitor, SKU 246808), 2,2′-Azobis(2-methylpropionitrile) (AIBN, SKU 441090), benzoin methyl ether (BME, 96%, SKU B8703), chlorotrimethylsilane (SKU 386529), phosphate buffered saline (PBS, pH 7.4, liquid, sterile filtered and suitable for cell culture, SKU 806552), and glycerol (SKY G9012) were purchased from Sigma Aldrich. Human Recombinant Keratinocyte Growth Factor (KGF) was purchased from Prospec, and the Alexa Fluor$^{TM}$ 488 Microscale Protein Labeling Kit was purchased from Invitrogen (catalog #A30006).

### A. Hydrogel blend preparation

0.5% crosslinked HEMA hydrogels were prepared by using HEMA, 0.5 vol % TMPTMA, and 0.2% BME dissolved in glycerol. The components were then mixed, degassed, and injected in between silanized glass slides that were separated by a 1.1 mm thick Teflon spacer, and allowed to polymerize under UV light for 30 minutes. HEMA/MMA and HEMA/MAA hydrogels were prepared using 5.89 mol % of MMA or 5.89 mol % MAA, 0.5 vol % TMPTMA, 94.1 mol % HEMA, 0.128 mol % AIBN. The components were then mixed and injected in between silanized glass slides that were separated by a 1.1 mm thick Teflon spacer, and were cured for 12 hours at 50ºC and then for 24 hours at 70ºC. All hydrogels were removed from the glass slides and washed three times at 70°C in triply distilled water. Glass slides were silanized in a dessicator under vacuum using chlorotrimethylsilane. Hydrogels were cut into 1x1 cm pieces and oven and vacuum dried prior to uptake, release, swelling, and ToF-SIMS experiments.

## B. FTIR-ATR measurements

Spectra of unlabeled KGF at the surfaces of the HEMA, HEMA/MMA, and HEMA/MAA hydrogels were acquired using a Spectrum Two™ FTIR Spectrometer equipped with a Universal ATR accessory containing a diamond/ZnSe crystal. 64 scans were acquired for each sample at 4 cm$^{-1}$ resolution. The Spectrum software was used to convert spectra from transmittance to absorbance, baseline correction, and ATR correction. Hydrogels were prepared by incubation in 250 nM KGF solutions in PBS for 2 hours, and spectra of hydrated samples were acquired. KGF spectra were determined by background subtraction using hydrogels that had been incubated in PBS only.

## C. Deconvolution of FTIR-ATR data

The amide I and amide I regions of the obtained spectra were Fourier self-deconvoluted using Origin Pro 2017. Parameters for deconvolution were 0.5 for gamma and 0.1 for smoothing.

## D. ToF-SIMS

1x1 cm hydrogels were incubated with 62.5 nM KGF in PBS for 2 hours. In order to remove salts and loosely bound KGF, all hydrogels were rinsed in stirred triply distilled water for two times for 30 seconds each. Hydrogels were then oven dried and vacuum dried. Hydrogels without KGF were also oven dried and vacuum dried. Positive secondary ion spectra were acquired on an ION-TOF V instrument (IONTOF, GmgH, Munster, Germany) using a Bi$_3^+$ primary ion source kept under static conditions (primary ion dose $< 10^{12}$ ions/cm$^2$). Three positive spectra from two samples per hydrogel blend type were collected from 100 x 100 μm regions (128 x 128 pixels). A pulsed flood gun was used for charge compensation. The ion beam was moved to a new spot on the sample after acquiring each spectrum. Spectra were acquired using high current bunched mode over a range of 0-1000 *m/z*. Mass resolution (*m/Δm*) was between 4000 and 7000. Hydrogel samples without KGF were calibrated using $CH_3^+$, $C_2H_3^+$, and $C_3H_5^+$, and hydrogel samples with KGF were calibrated using $CH_3^+$, $C_2H_3^+$, $C_3H_5^+$, and $C_5H_{10}N^+$. Mass calibration errors were kept below 20 ppm.

### E. *Multivariate Analysis*

Peak lists were compiled for PCA analysis using the Ion-Tof SurfaceLab software. PCA analysis of the hydrogel blends without KGF used a peak list consisting of all peaks in the 0-200 *m/z* range greater than 3 times the background. PCA analysis of the hydrogel blends with KGF used a peak list of reported amino acid fragments in the 0-200 *m/z* range. Amino acid fragments that overlapped with fragments from HEMA, MMA, or MAA were eliminated from the peak list, and resulted in a final list of 11 amino acids. Peak lists were imported into the NESAC/BIO NBToolbox "spectragui" program written for MATLAB (Mathworks, Inc. Natick, MA)[40]. Data sets were normalized using the sum of selected peaks, mean centered, and square root transformed prior to PCA analysis.

### F. *Fluorescence monitoring of KGF uptake and release from hydrogel blends*

KGF was labeled with Alexa Fluor™ 488 using the Microscale Protein Labeling Kit. Fluorescence measurements were taken using a Qubit 3.0 Fluorometer. Uptake and release into and from the blends was monitored by preparing cuvettes containing 1 mL of labeled KGF solution in PBS.  1x1 cm oven-dried hydrogels were placed in the solution and kept in the dark, and the decrease in fluorescence of the labeled KGF was monitored over 24 hours.  Uptake fluorescence was corrected for water content taken up into the hydrogel over time (see swelling experiment).  Release fluorescence measurements were made by placing the protein loaded hydrogel in a cuvette containing 1 mL fresh PBS in a 37°C water bath and monitoring the increase in fluorescence over 2 hours.  2 replicates were performed for each hydrogel type.  Release assays were limited to 2 hours given that previous (not reported here) experiments show that all possible release from HEMA based hydrogels occurs within this time frame.

## *G.   Swelling experiments*

1x1 cm oven and vacuum dried HEMA, HEMA/MMA, and HEMA/MAA hydrogels were used for swelling experiments.  Hydrogels were initially weighed, and then placed into cuvettes containing 1 mL PBS.  The increase in weight of the hydrogels was monitored at 5 mins, 15 mins, 30 mins, 1 hours, and 2 hours.  2 replicates were performed for each hydrogel type.  Percentage swelling was calculated using the equation $(M_t - M_o)/M_o \times 100\%$, where $M_t$ is the mass of the swelling hydrogel at time t, and $M_o$ is the initial mass of the dry hydrogel.

## III. RESULTS

### A. *Surface characterization of the hydrogel blends by ToF-SIMS and PCA*

The surface percent of the 5.89 mol % MAA in HEMA and 5.89 mol % MMA in HEMA blends was calculated by using the ($C_4H_9^+$/($C_4H_9^+$ + $C_2H_5O^+$)) and the ($C_2H_3O_2^+$/($C_2H_3O_2^+$ + $C_2H_5O^+$)) peak area ratios respectively[41, 42]. The surface percent of MAA in the HEMA/MAA blend is 5.39 ± 1.98 % and the surface percent of MMA in the HEMA/MMA blend is 2.84 ± 0.51 %. The large standard deviation in the surface percent of MAA in the HEMA/MAA blend indicates the possibility of phase segregation at the surface of MAA regions at the surface, while this is not seen in the HEMA/MMA blend. The distribution of MAA and MMA surface percent across the samples and representative spectra of the three different blends are shown in figures S1-1 and S1-2 of the supplementary material.

PCA on these blends utilizing a peak list of all the peak contributions from HEMA between 12-200 m/z was performed in order to determine whether the surfaces were distinguishable. PC1 captured 61% of the variance in the data set, and the HEMA/MAA and HEMA/MMA surfaces were distinguishable at the 95% confidence level[43], while the HEMA/MAA and HEMA surfaces were not distinguishable, and the HEMA/MMA and HEMA were distinguishable. PC1 loadings showed that the HEMA/MMA blend was distinguished from the HEMA/MAA and HEMA blends by $C_2H_6^+$, $C_3H_8O^+$, $C_5H_{10}^+$, and $C_4H_9O^+$, while the HEMA/MAA and HEMA blends are distinguished from the MMA blend by $C_4H_{10}^+$ and $C_4H_{11}^+$. The PCA scores also show that the HEMA/MMA samples appear relatively homogeneous while heterogeneity was seen in the HEMA/MAA and HEMA blends.  PC2 captures 19% of the variance and the three blends are not distinguishable at the 95% confidence interval as seen in figure S1-3 of the supplementary material.  However, the HEMA/MAA blend exhibits both positive and negative scores while the HEMA/MMA and HEMA blends only exhibit positive scores.  The peak with the most prominent negative loading in PC2 was due to $C_2H_5O^+$.

While the HEMA/MAA and HEMA blend surfaces are not distinguishable by PCA analysis, differences in protein orientation and conformation at the surfaces of these blends is still of interest.  Additional factors such as phase segregation and topography may contribute to differences in protein orientation and conformation at the surface, and are the subject of future investigations.

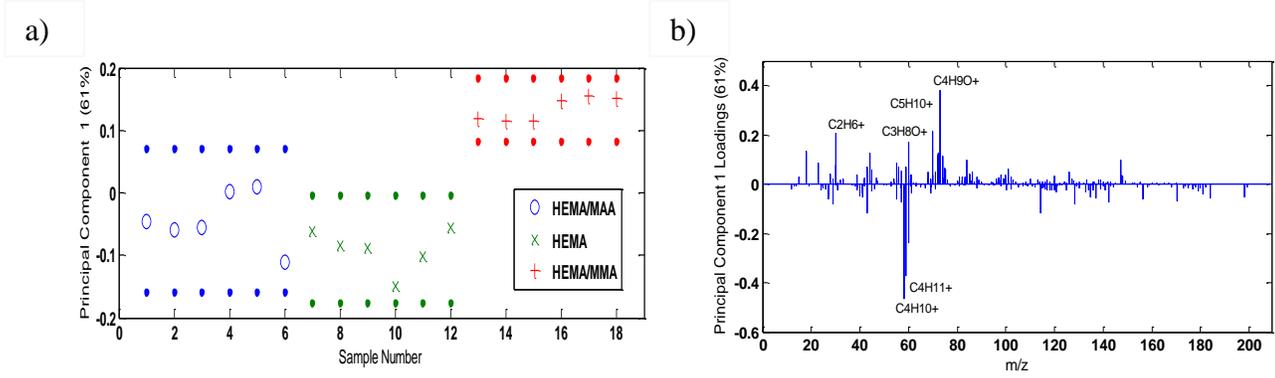

Figure 1. (a) PC1 confidence limits (95%) of the three hydrogel blend surfaces and (b) PC1 loadings

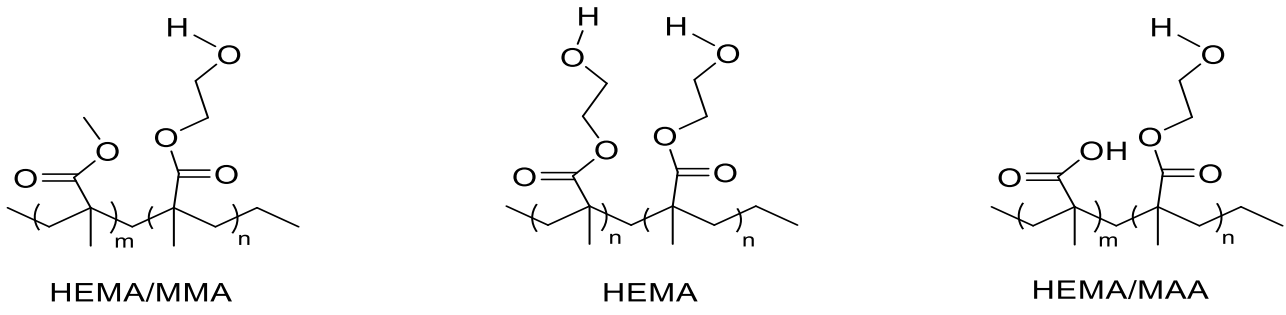

Figure 2. Structures of the hydrogel blends that were crosslinked for the formation of pores

## B. KGF conformation: KGF at the surface of the HEMA/MAA blend appears to be denatured; FSD analysis

Figure 3 shows the amide I and amide II regions of FTIR-ATR spectra corresponding to KGF conformation at the three different hydrogel surfaces. The highest concentration of KGF is seen at the surface of the HEMA/MAA blend, potentially due to increased protein diffusion at the surface and into the porous network due to additional hydrogen bonding from the carboxylic acid in MAA. The HEMA-only hydrogel shows the lowest KGF surface concentration. The HEMA/MMA blend may present a higher KGF surface concentration than the HEMA blend due to increased adsorption on the surface.

Fourier self-deconvolution of the amide I regions has been used to interpret conformational changes in KGF. The amide I region is our region of focus due to its ability to reflect changes in protein secondary structure. The relative intensities of the peaks at 1634 and 1643 cm$^{-1}$ peaks corresponding to the extended strands and irregular/disordered region have been shown in our prior work on KGF conformation to be indicative regarding whether KGF is folded in its native conformation or its heat-denatured conformation. Our previous work has shown that equal intensities indicate a native conformation while unequal intensities where the intensity of the 1643 cm$^{-1}$ peak is lower than the intensity of the 1634 cm$^{-1}$ peak indicate a denatured conformation[11]. A denatured conformation of KGF is seen at the surface of the HEMA/MAA blend while a native conformation is seen at the surface of the HEMA/MMA and HEMA blends. Our previous work has also shown that extended interactions with the HEMA hydrogel result in eventual denaturation of KGF due interactions mimicking hydrogen-bond driven KGF-heparin interactions[11]. A denatured conformation after only a two-hour incubation at the HEMA/MAA surface indicates that these interactions may be occurring at a much faster rate due to the presence of MAA-based hydrogen bonding. Furthermore, a much lower relative intensity in comparison to the peaks at 1634 and 1643 cm$^{-1}$ peaks is seen at 1651 cm$^{-1}$, which corresponds to the loops of KGF (Figure 6). There are two loops in KGF; the receptor (heparin)-binding loop and the loop at the beginning of the protein that is right next to an additional heparin-binding site. Therefore, it is possible that the HEMA/MAA blend is disrupting the loops through KGF-heparin mimicking interactions, resulting in a denatured conformation. The HEMA/MMA blend alternatively appears to target hydrophobic regions of KGF. For example, a peak at 1634 cm$^{-1}$ corresponding to the extended strands region between 1622-1634 cm$^{-1}$ disappears in the KGF spectrum, and the crystal structure of KGF shows that the extended strands create a tightly bound hydrophobic core (Figure 6). Overall, KGF adopts a native conformation on the surfaces of the HEMA and HEMA/MMA blends after a two-hour incubation, while it appears to denature at the surface of the HEMA/MAA blend.

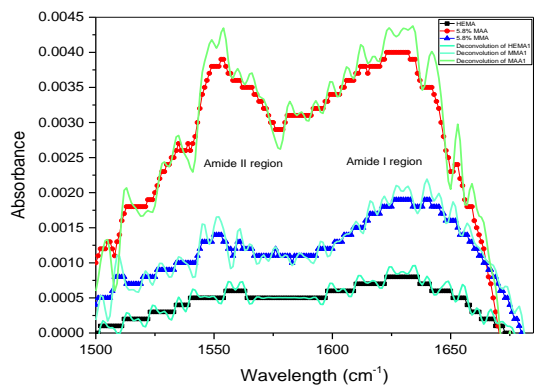

Figure 3. Original and Fourier self-deconvoluted spectra of KGF on the surfaces of the three hydrogel blends.

## C. KGF orientation: ToF-SIMS and PCA show that KGF hydrophobic amino acids are exposed at the HEMA/MAA blend surface while hydrophilic amino acids are exposed at the HEMA/MMA blend

Differences in KGF orientation at the surface of the hydrogel blends due to MMA and MAA addition were studied by PCA of the ToF-SIMS results. Differences in orientation were best revealed when a peak list consisting of only amino acids were used. Positive ToF-SIMS spectra were used for PCA analysis due to a history of positive spectra being informative regarding protein orientation[13, 14, 16, 18, 43]. Amino acid fragments that overlapped with peaks from HEMA, MMA, or MAA were omitted from the peak list, which resulted in a final peak list of 11 amino acids as shown in figure S2-2 of the supplementary material.

Figure 4 of PC1 scores shows that PC1 captures 54% of the variance in the data. KGF orientation at the HEMA/MMA and HEMA/MAA blend surface regions are distinguishable at the 95% confidence level[43]. KGF orientation at the HEMA surface regions displays characteristics seen in both the HEMA/MMA and HEMA/MAA blends.

Amino acids used in the peak list were characterized as polar/hydrophilic and aliphatic/hydrophobic in order to determine whether the PC1 loadings indicated a pattern for KGF orientation. Hydrophilic amino acids were found to have positive loadings while hydrophobic amino acids had negative loadings. PC1 loadings shown in Figure 4 indicate that the aliphatic amino acids alanine and isoleucine/leucine were detected in high intensities at the surface of the HEMA/MAA blend while polar amino acids serine, threonine, glutamine, arginine, and phenylalanine were detected in high intensities at the surface of the HEMA/MMA blend. These results suggest that aliphatic amino acids are oriented outwards while polar and mostly hydrophilic amino acids interact with the HEMA/MAA surface, and that polar/hydrophilic amino acids are oriented outwards while aliphatic amino acids interact with the HEMA/MMA surface. Observations such as these on PMMA-containing surfaces have been previously reported[44].

PC2 captures 37% of the variance and creates a distinction between the 2 different HEMA/MMA samples that were studied (3 different regions studied on each sample, see figure S2-1 in the supplementary material). PC2 scores indicate that there is significant heterogeneity in protein orientation among the HEMA/MMA samples. This suggests the possibility of different amounts of phase segregation at the surface caused by the formulation method that may be causing minor differences in protein orientation among the HEMA/MMA samples. The HEMA/MAA and HEMA blends appear homogenous in PC2, and PC2 loadings indicate no clear pattern distinguishing orientation at the surface of the HEMA/MMA blend.

Figure 6 is a structure of KGF with hydrophobic (pink, Ala, Gly, Val, Ile, Leu) and hydrophilic (blue, all remaining) amino acids labeled and shows that aliphatic/hydrophobic amino acids are mostly found to flank the beginning and end of the 6 beta sheets near the solvent exposed reverse turns, while the polar/hydrophilic amino acids are found in the middle of the beta sheets. Though only 11 (12 including Ile/Leu) amino acids are included in the PCA, we believe that our results are representative of the system because Figure 6 shows that hydrophobic and hydrophilic amino acids tend to localize separately in KGF. Given this knowledge, we believe that inclusion of the remaining amino acids in the PCA would be beneficial but is not possible due to the fragmentation of the hydrogels, and we are comfortable forming the conclusions discussed below.

Of particular importance are the positive loadings of serine, and threonine in PC1. Mutations in residues S122, and T126 to alanine correspond to the biological activity being reduced to 70%, and 60% when evaluated based on tritiated thymidine uptake in Balb/MK cells[45]. These residues are in the receptor-binding loop (residues 122-132). Their higher normalized intensities at the HEMA/MMA surface seen in Figure 5 in comparison to the HEMA/MAA surface indicate an orientation more likely to bind the KGF receptor.

The results suggest that the HEMA/MAA blend specifically targets aliphatic/hydrophobic amino acids of the beta sheets near solvent exposed regions and does not show accessibility of residues involved in receptor binding, while the HEMA/MMA blend specifically targets polar/hydrophilic amino acids found within the receptor binding loop and beta sheets. Implications of these differing orientations on the receptor binding abilities of KGF and the biological activity of these hydrogels on wound closure are the subject of future experiments.

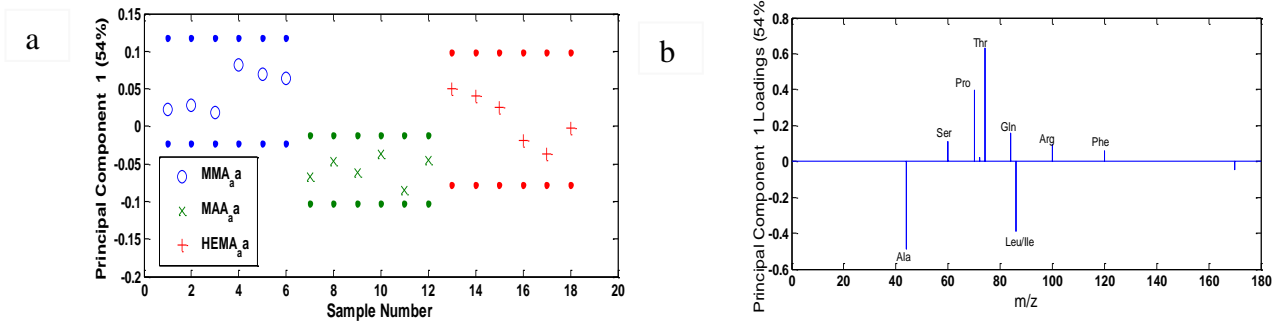

Figure 4. (a) PC1 confidence limits of protein orientation on the hydrogel surfaces and (b) PC1 loadings

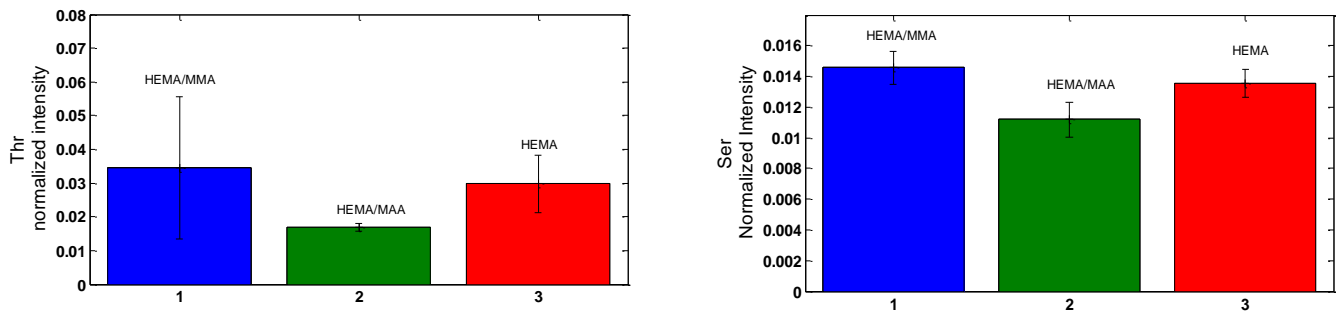

Figure 5. Peak intensities of serine and threonine when normalized by the sum of selected peaks

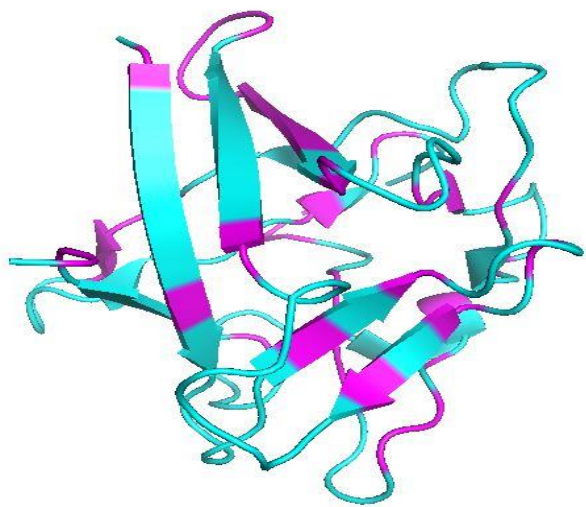

Figure 6. Crystal structure of KGF (PDB ID: 1QQK[46]) with hydrophobic (pink) and hydrophilic amino acids labeled (blue). Structure created in PyMol.

## D. Effect of hydrogel composition on bulk properties

Swelling, uptake, and release of Alexa Fluor™ 488-labeled KGF (AF-KGF) from the hydrogel blends was evaluated in order to determine the effects of the changes in bulk and surface composition on water content and KGF delivery. Surprisingly, and contrary to previously published literature[34], changes in composition had no statistically significant effect on swelling over 2 hours (at the 0.05 level, $F_{2,2} = 3.71$, $p > 0.05$) with 45, 50, and 52% swelling for the HEMA/MMA, HEMA/MAA, and HEMA blends respectively. The percent of initially loaded AF-KGF taken up into the hydrogels after 24 hours also was not statistically significant (at the 0.05 level, $F_{2,2} = 2.40$, $p > 0.05$) with 7, 11, and 20% uptake for the HEMA/MMA, HEMA/MAA, and HEMA blends respectively. Finally, the percent of loaded AF-KGF released after 2 hours was not statistically significant either (at the 0.05 level, $F_{2,2} = 0.96$, $p > 0.05$) with 62, 58, and 44% of loaded AF-KGF released. Therefore, the changes in bulk and surface composition of the hydrogels upon addition of MMA and MAA did not improve our previously reported concerns regarding incomplete release of KGF potentially due to trapping and denaturation of KGF at the hydrogel surface over time. It is possible that the addition of larger percentages of MMA or MAA may address these concerns. These results highlight the importance of evaluating both surface and bulk properties of newly modified materials when evaluating drug delivery systems. Complete swelling, uptake, and release data can be found in figure S3-1 of the supplementary file. The HEMA/MAA and HEMA/MMA blends differentially influence protein orientation and conformation without influencing swelling or protein uptake and release. The three different KGF loaded hydrogel blends may therefore vary in biological activity, based on previous reports on the influence of conformation and orientation on biological activity. This will be the subject of future research efforts.

## IV. DISCUSSION

The results of this study show that blends of HEMA hydrogels made to promote adsorption of protein at the surface or diffusion of protein into the surface affect protein orientation and conformation differentially. We show that KGF detected at the surface of the HEMA/MMA hydrogel has a (1) greater percent of its amino acids oriented away from the surface than KGF at the surface of the HEMA/MAA hydrogel, with its hydrophobic amino acids interacting with the surface and it hydrophilic amino acids oriented outward, (2) KGF in an orientation likely to bind the KGF receptor, and that (3) KGF secondary structure remains in its native conformation. Our observations regarding the orientation of hydrophobic and hydrophilic amino acids have been previously observed in AFM studies of BSA adsorbed at the surface of MMA/AA (acrylic acid) block copolymers[44]. Hydrophilic groups of BSA were oriented away from the MMA surface and this was detected by differences in adhesive force between BSA hydrophilic/hydrophobic groups and the AFM tip. The high intensities of PC loadings of amino acids responsible for biological activity have also been previously reported to be indicative of receptor-binding orientations[16]. HEMA/MMA swelling and KGF release profiles in comparison to HEMA are unaffected by these surface-level changes.

In contrast, we show that KGF detected at the surface of the HEMA/MAA hydrogel has (1) a smaller percent of its amino acids oriented away from the surface with its hydrophobic amino acids oriented away from the surface, (2) KGF in an orientation that doesn't expose amino acids involved in receptor binding, and (3) KGF in a conformation that resembles that of denatured KGF.  HEMA/MAA swelling and KGF release profiles in comparison to HEMA are also unaffected by these surface-level changes.

PCA of the hydrogel blend surfaces (Figure 1) indicates that the HEMA/MMA and HEMA/MAA blends are distinguishable surfaces while the HEMA/MAA and HEMA surfaces are not.  PCA as well as characterization of surface content of MAA or MMA in the blends shows that the HEMA/MMA blend is relatively homogeneous while the HEMA/MAA blend is heterogeneous.  Surprisingly, KGF orientation at the HEMA/MMA blend is heterogeneous while KGF orientation HEMA/MAA blend is homogeneous (Figure 4).  This suggests that contributions from phase segregation and topography at these hydrogel blend surfaces and their effects on KGF orientation and localization are of interest in follow-up studies using SIMS imaging, XPS, and NEXAFS.  PCA analysis of KGF orientation on the three surfaces shows that orientation on the HEMA/MMA blend is distinguishable from the orientation on the HEMA/MAA blend while the orientation on the HEMA blend cannot be distinguished from the other orientations on the other two blend surfaces, and has loadings in PC1 that overlap with loadings contributing to orientation on HEMA/MMA and HEMA/MAA.   Therefore, analysis of KGF orientation and characterization of the hydrogel blend surfaces tells us that differences in between the blends are specifically due to the presence of the additives MMA or MAA.

The PCA results of KGF orientation suggest that the hydrophobic amino acids of KGF including alanine, isoleucine, leucine are oriented outwards facing away from the HEMA/MAA surface while the hydrophilic amino acids likely interact with the hydrophilic HEMA/MAA surface (Figure 4). The crystal structure of KGF indicates that these hydrophobic amino acids are mostly found in regions flanking the beginning and ends of the 5 extended strands and 6$^{th}$ disordered extended strand. These amino acids are therefore in close proximity to the solvent exposed reverse turns of KGF as seen in Figure 6. Amino acids such as serine, and threonine are not seen in PC loadings corresponding to KGF orientation at the HEMA/MAA surface. The FTIR-ATR spectrum of KGF at the HEMA/MAA surface indicates disruption of the extended strands and disordered regions leading to the adaption of a conformation that resembles that of previously reported heat-denatured KGF. It is likely that this observed conformation with changes in the extended strands and disordered regions is caused by the distortion of KGF due to the flanking hydrophobic regions of the extended strands being oriented away from the HEMA/MAA surface while the interior hydrophilic regions of the extended strands are interacting with the HEMA/MAA surface.

The FTIR-ATR spectrum also shows disruption of the loop regions. Past work on KGF conformational changes at the surface of the HEMA hydrogel over time has shown that the sequence of events leading eventual adoption of a conformation resembling that of heat-denatured KGF is initiated by an interaction between HEMA and the loops that potentially mimics the hydrogen bond-mediated interaction between KGF and its ligand heparin[11]. Given that the PCA shows that KGF orientation at the HEMA surface and KGF orientation at the HEMA/MAA surface are not distinguishable at the 95% confidence level, it is possible that a similar conformational mechanism involving a loop-surface interaction is leading to the denatured conformation of KGF at the HEMA/MAA surface. While such a mechanism can only be proven by follow-up studies of 2D correlation spectroscopy of KGF conformation at the HEMA/MAA surface, it is likely given (1) the structural similarity between MAA and HEMA with both containing hydrogen-bonding capability, and (2) PCA results of the blends in Figure 1 showing that the HEMA/MAA and HEMA blends are not distinguishable at the 95% confidence level in PC1 and PC2.

PCA results of the amino acids at the HEMA/MMA surface shows that hydrophilic/polar amino acids serine, proline, threonine, glutamate, arginine, and phenylalanine are oriented outwards while it is likely that hydrophobic amino acids are interacting with the HEMA/MMA surface in hydrophobic-hydrophobic interactions. Serine, and threonine are found in the receptor binding loop, and display high normalized intensities at the HEMA/MMA in comparison to the other two surfaces. Interestingly, hydrophilic/polar amino acids are found both (1) buried within the extended strands away from the solvent exposed reverse turns and (2) also in the solvent exposed loops and disordered regions. This suggests that both the hydrophobic nature of MMA and hydrophilic nature of HEMA may be contributing to interactions with different parts of KGF. This is backed up by the fact that the orientation of KGF at the HEMA surface is not distinguishable from the orientation of KGF at the HEMA/MMA surface at the 95% confidence level, even though KGF orientation at the HEMA/MAA and HEMA/MMA surfaces is distinguishable.

The FTIR-ATR spectrum of KGF at the HEMA/MMA surface indicates that the extended strands are disrupted, but not in a way that leads to disruption of the disordered regions or loops which can lead to denaturation seen at the HEMA/MAA surface.

Overall, in comparison to the orientation of KGF seen at the HEMA/MAA surface, more regions of KGF are accessible and oriented outwards as indicated by the PCA results.

The purpose of this study has been to characterize how the addition of MAA or MMA to the HEMA blend affects KGF orientation differentially, while also characterizing whether KGF conformation remains intact or appears to denature. Given that the results of our previous study showed that KGF denatured after extended time spent at the HEMA surface during the release process, we hypothesized that MAA addition may contribute to potential denaturation of KGF due to additional hydrogen-bonding capability, while MMA may lead to a decreased interaction with KGF due to its hydrophobic nature. Our results show that likely due to the additional hydrogen-bonding capability of the HEMA/MAA blend, the surface concentration of KGF is higher than HEMA/MMA and HEMA as shown by the area under the amide I region in the FTIR-ATR spectrum (Figure 3), even though the overall percent of KGF taken up into the three hydrogels is indistinguishable by one-way ANOVA (figure S3-1 in the supplementary material). Although the HEMA/MMA blend has a lower KGF surface concentration, KGF maintains a close to native conformation, as shown in the FTIR-ATR spectrum, and allows for both internal amino acids of the extended strands and large portions of solvent exposed regions of KGF to be oriented away from the HEMA/MMA surface. In contrast, the HEMA/MAA blend leads to potential denaturation of KGF.

It is important to note that surface concentration indicated by the FTIR-ATR spectra provides depth resolution of 1-2 µm into the surface, while the ToF-SIMS provides depth resolution of 1-2 nm of the surface. Therefore, to summarize our interpretation of the results of this study, we believe that the HEMA/MAA blend possesses characteristics that are beneficial in obtaining a high surface concentration of KGF. The hydrogen-bonding capability at the surface likely leads to diffusion of KGF into the porous network and adsorption at the surface, which is observed in the high surface concentration of KGF seen at the HEMA/MAA surface. However, this strong interaction likely leads to a less accessible orientation of KGF at the surface with receptor binding amino acids facing towards the hydrogel surface. The ability of the HEMA/MAA blend to act as a drug delivery vehicle therefore depends on the strength of the HEMA/MAA interaction with KGF in comparison to the strength of the KGF-receptor interaction in the presence of epithelial cells containing the KGF receptor. We believe that the interaction between HEMA/MAA and KGF is potentially too strong, which leads to KGF denaturation and incomplete release. However, this hypothesis can only be tested by follow up experiments that test the ability of KGF at the HEMA/MAA surface to bind its receptor in either *in vitro* heparin binding studies, or studies of the efficiency of wound closure in *in vivo* cellular assays that have been previously developed in our labs.

Alternatively, we believe that the HEMA/MMA blend has a weaker interaction with KGF that is reflected by the lower surface concentration seen in the FTIR-ATR spectrum. However, the overall concentration of KGF taken up into the HEMA/MMA hydrogel is indistinguishable from the KGF uptake in the other blends. The HEMA/MMA blend doesn't disrupt the conformation of KGF and denaturation is avoided potentially due to these weaker protein-material interactions, and the PCA results suggest that more regions of KGF are oriented away from the HEMA/MMA surface, and are more receptor-accessible. This interaction is possibly beneficial in the development of the ideal drug delivery device even though the HEMA/MMA bulk properties in the context of swelling, and uptake/release properties appear similar to the HEMA/MAA and HEMA blend.

## V. SUMMARY AND CONCLUSIONS

We have characterized the abilities of HEMA/MMA and HEMA/MAA hydrogels to promote adsorption and diffusion at their surface related to KGF orientation and conformation by combining ToF-SIMS/PCA of the hydrogels, ToF-SIMS/PCA of KGF orientation, and FTIR-ATR spectra of KGF conformation. While neither PMAA nor PMMA are ideal drug delivery vehicles, this approach has allowed for the characterization of differences in KGF orientation/conformation caused by new properties of the modified HEMA surfaces. The HEMA/MMA formulation allows for a more receptor-accessible orientation of KGF that will potentially result in higher biological efficacy in expediting wound closure. The HEMA/MAA formulation has an increased surface concentration of KGF, but the strength of the hydrogen bonding interactions between KGF and the hydrogel lead to a less receptor-accessible orientation and denaturation. We believe that differences in receptor binding and efficacy in wound closure among the hydrogels will likely be due to differences in KGF orientation and conformation instead of differences in swelling and KGF release profiles. We aim to focus future studies defining the role of phase segregation and porous topography on KGF localization and orientation/conformation, and we will evaluate the effects of KGF orientation, conformation, and localization on receptor binding and wound closure through *in vitro* heparin binding assays and previously developed *in vivo* wound closure assays[7].

## ACKNOWLEDGMENTS

The authors would like to thank undergraduate student Zoe Vaughn in her support efforts in hydrogel synthesis as well as the members of the Gardella group for thoughtful discussions. These studies were supported by funding from the John and Frances Larkin Endowment awarded to J.A.G. The authors also thank Dan Graham, Ph.D., for developing the NESAC/BIO Toolbox used in this study and NIH grant EB-002027 for supporting the toolbox development.

# Differential orientation and conformation of Keratinocyte Growth Factor at HEMA, HEMA/MMA, and HEMA/MAA observed at hydrogel surfaces developed for wound healing

Shohini Sen-Britain[a], Wesley Hicks[b], Robert Hard[c], Joseph A. Gardella Jr.[a]

[a] State University of New York at Buffalo, Department of Chemistry, 475 Natural Sciences Complex, Buffalo, NY 14221, USA

[b] Roswell Comprehensive Cancer Center, Department of Head and Neck/Plastic and Reconstructive Surgery, 665 Elm Street, Buffalo, NY 14203, USA

[c] State University of New York at Buffalo, Jacobs School of Medicine and Biomedical Sciences, Department of Pathological and Anatomical Sciences, 955 Main St, Buffalo, NY 14203, USA

**Supporting Information**

**S1: Surface Characterization results of HEMA/MMA, HEMA, and HEMA/MAA hydrogel blends**

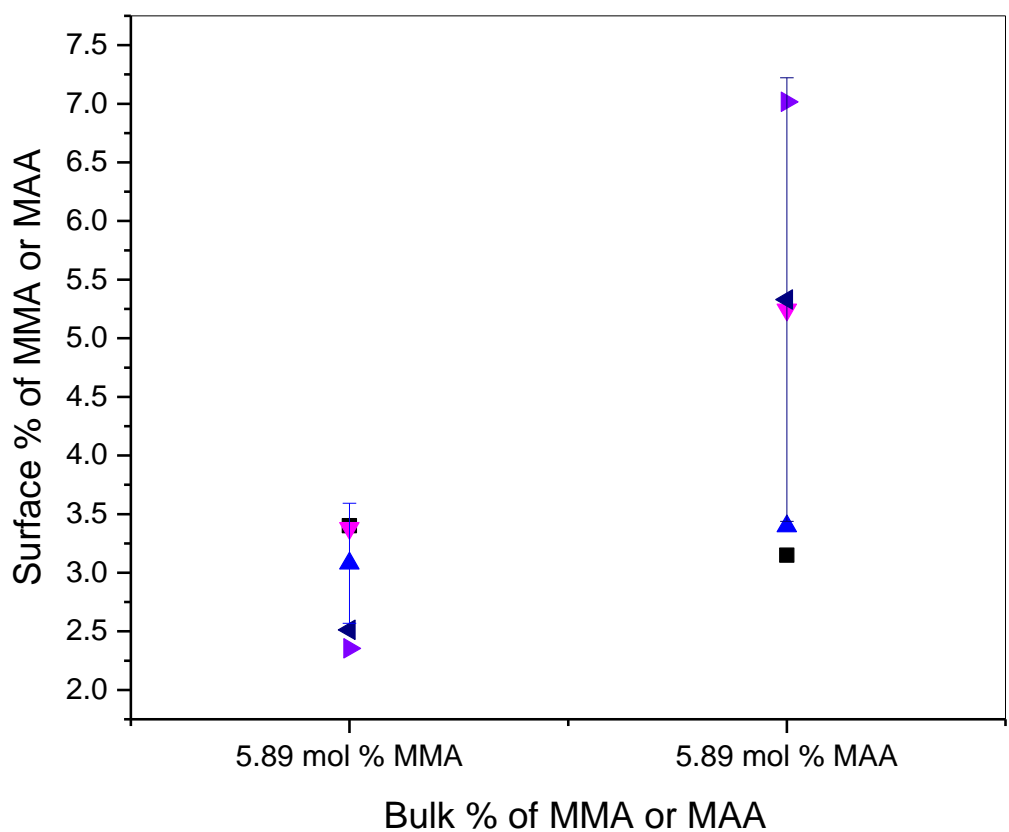

**S1-1**. Surface vs. bulk % of MAA and MMA for HEMA/MAA and HEMA/MMA blend surface regions with error bars calculated by ($C_4H_9^+/(C_4H_9^+ + C_2H_5O^+)$) and the ($C_2H_3O_2^+/(C_2H_3O_2^+ + C_2H_5O^+)$) peak area ratios from the ToF-SIMS spectra respectively

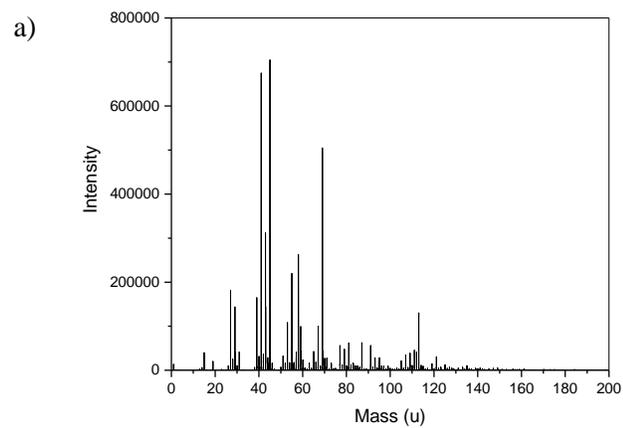

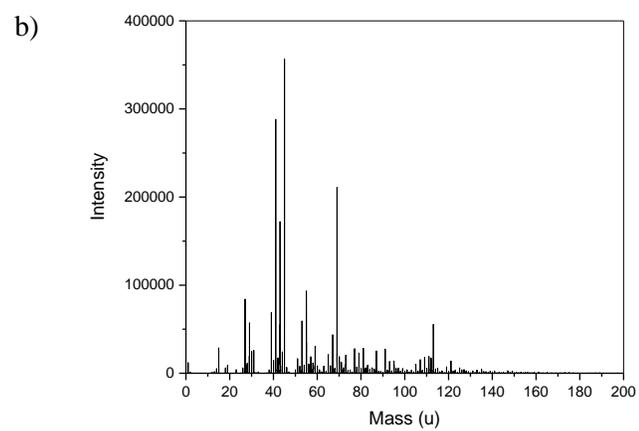

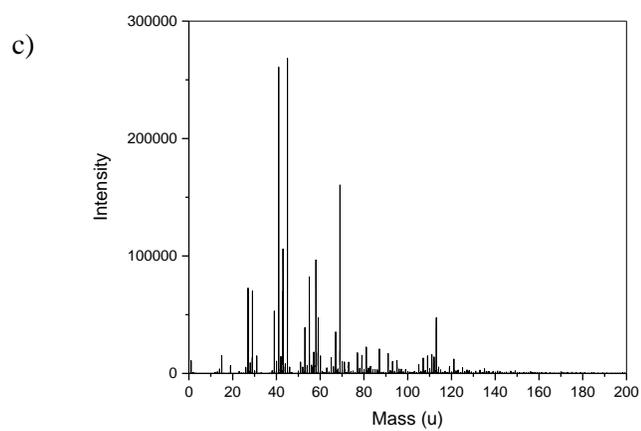

**S1-2**. Representative ToF-SIMS spectra of (a) 5.89 mol % MAA in HEMA (b) 5.89 mol % MMA in HEMA (c) HEMA hydrogel surfaces

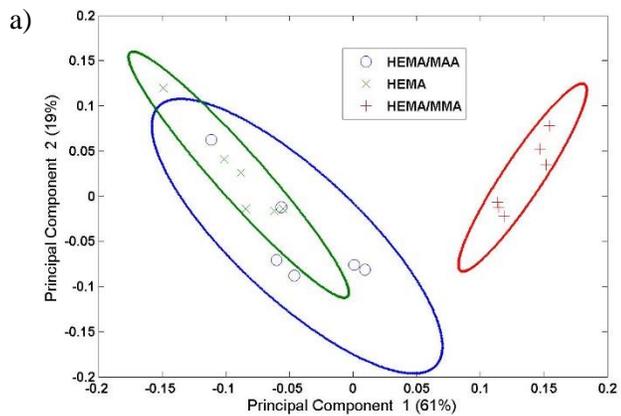 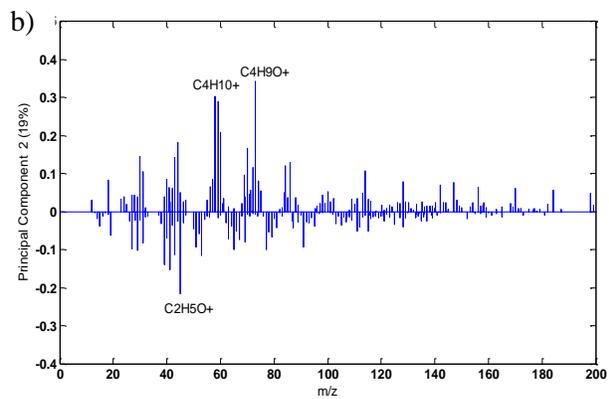

**S1-3:** (a) PC2 vs. PC1 of the three hydrogel blends and (b) PC2 loadings for the three hydrogel blends

## S2: PCA of KGF orientation at the surface regions of the hydrogel blends

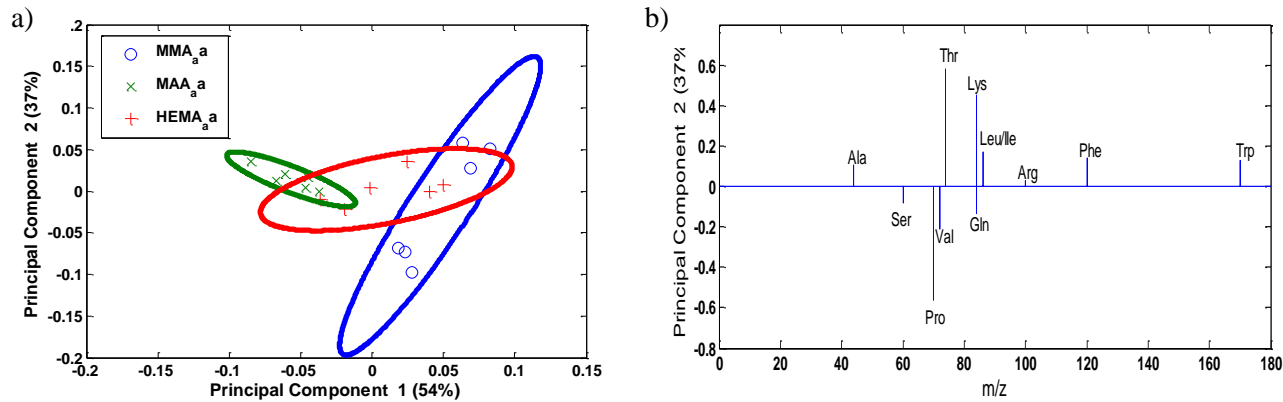

**S2-1**. (a) PC2 v. PC1 of KGF orientation at the hydrogel surfaces and (b) PC2 loadings

| Amino Acid | m/z |
|---|---|
| Alanine ($C_2H_6N+$) | 44.05 |
| Serine ($C_2H_6NO+$) | 60.0425 |
| Proline ($C_4H_6N+$) | 70.0723 |
| Valine ($C_4H_{10}N+$) | 72.065 |
| Threonine ($C_3H_8NO+$) | 74.0606 |
| Glutamine ($C_4H_8NO_2+$) | 84.0411 |
| Lysine ($C_5H_{10}N+$) | 84.0836 |
| Leucine/Isoleucine ($C_5H_{12}N+$) | 86.098 |
| Arginine ($C_4H_{10}N_3+$) | 100.08 |
| Phenylalanine ($C_8H_{10}N+$) | 120.0802 |
| Tryptophan ($C_{11}H_8NO+$) | 170.057 |

**S2-2:** Amino acids included in peak list for positive spectra used in PCA

**S3: Bulk characterization of swelling, uptake, and release of AF-KGF from the HEMA, HEMA/MMA, and HEMA/MAA hydrogels**

| Blend | Swelling % of total weight taken up of water | Uptake % of initial AF 488-KGF soln. loaded | Release % of loaded conc. released |
|---|---|---|---|
| MMA in HEMA | 45.16 ± 0.17 | 6.52 ± 1.80 | 61.92 |
| MAA in HEMA | 50.04 ± 2.94 | 11.13 ± 0.01 | 57.85 |
| HEMA | 52.48 ± 3.71 | 20.02 ± 8.29 | 43.66 ± 11.90 |

**S3-1**. Values for percentage swelling after 2 hours, percent of 15 nM AF-KGF taken up into hydrogels after 24 hours, and percent of final loaded AF-KGF released after 2 hours. Replicates for HEMA/MMA and HEMA/MAA release are not available.